\documentclass[prl,aps,amssymb,twocolumn,showpacs,floatfix]{revtex4}
\usepackage[dvips]{graphicx}
\def\be{\begin{equation}}
\def\ee{\end{equation}}
\def\bea{\begin{eqnarray}}
\def\eea{\end{eqnarray}}
\def\ve{\varepsilon}

\def\bE{{\bf E}}

\def\wc{\omega_c}
\def\ve{\varepsilon}

\def\w{\omega}
\def\St{{\rm St}}

\begin{document}

\title{Theory of the fractional microwave-induced resistance oscillations}
\author{I.A.~Dmitriev$^{1,*}$}
\author{A.D.~Mirlin$^{1,2,\dagger}$}
\author{D.G.~Polyakov$^{1}$}
\affiliation{$^{1}$Institut f\"ur Nanotechnologie, Forschungszentrum
Karlsruhe, 76021 Karlsruhe, Germany\\
$^{2}$Institut f\"ur Theorie der kondensierten Materie,
Universit\"at Karlsruhe, 76128 Karlsruhe, Germany}

\date{\today}

\begin{abstract} We develop a systematic theory of
microwave-induced oscillations in magnetoresistivity of a 2D electron gas
in the vicinity of fractional harmonics of the cyclotron resonance, observed in
recent experiments. We show that in the limit of well-separated Landau levels
the effect is dominated by a change of the distribution function induced by
multiphoton processes. At moderate magnetic
field, a single-photon mechanism originating from the microwave-induced
sidebands in the density of states of disorder-broadened Landau levels becomes
important.
\end{abstract}


\maketitle

\noindent
Recently, a number of remarkable nonequilibrium phenomena have
been discovered in a 2D electron gas (2DEG) under strong ac and dc
excitation. Most attention has been attracted to the
microwave-induced resistivity oscillations
(MIRO) \cite{zudov01}, particularly following the spectacular
observation of ``zero-resistance states''(ZRS) in the minima of
the oscillations \cite{mani02,zudov03}.  Two mechanisms of the MIRO
were proposed (``displacement'' \cite{ryzhii,durst03,VA} and
``inelastic'' \cite{dmitriev03,long}), in both
of which the MIRO originate from the oscillatory density of states
(DOS) $\nu(\ve)$ of disorder-broadened Landau levels (LLs).
Both mechanisms reproduce the observed
phase of the $\w/\wc$--oscillations ($\w$ and $\wc=eB/mc$ are
the microwave and cyclotron frequencies, respectively).  The
displacement mechanism \cite{ryzhii,durst03,VA} accounts for
microwave-assisted scattering off disorder in the presence of dc electric field and
produces temperature-independent MIRO, in disagreement with the experiments.
By contrast, the inelastic mechanism \cite{dmitriev03,long} is related to
the microwave-induced oscillatory changes in the energy distribution
of electrons and yields the MIRO with an amplitude
proportional to the inelastic scattering time $\tau_{\rm in}\propto T^{-2}$.  At
relevant $T\sim 1$~K, the inelastic
mechanism dominates and the corresponding theory \cite{long}
reproduces the experimental findings \cite{zudov01,mani02,zudov03}.

Remarkably, in addition to the peak-valley structure near integer $\w/\wc$
(``integer MIRO'', or IMIRO), later
experiments at elevated microwave
power \cite{multi,dorozhkin06} reported similar features near certain fractional
values, $\w/\wc=1/2,\,3/2,\,5/2,\,2/3$ (``fractional
MIRO'', or
FMIRO), as well as ``fractional'' ZRS
\cite{multi}
(less pronounced FMIRO
features were also observed earlier \cite{zudov04}). It was
proposed that the FMIRO can be explained in terms of a
multiphoton displacement mechanism \cite{leiliumulti} or, within the framework
of the inelastic mechanism \cite{dmitriev03,long}, in terms of a series of
multiple single-photon transitions \cite{dorozhkin06,crossover}.

In this Letter, we develop a systematic theory of the fractional MIRO. We
demonstrate that in the limit of well-separated LLs the FMIRO are
dominated by the multiphoton inelastic mechanism, while at weaker magnetic
field a microwave-induced spectral reconstruction (MISR) provides a
competing single-photon contribution.  Both these mechanisms were
disregarded previously. Similarly to the IMIRO, in the fractional case
the multiphoton displacement mechanism \cite{leiliumulti} only gives a
parametrically smaller contribution. As far as the mechanism
\cite{dorozhkin06,crossover} is concerned, it is
effective only close to the magnetic field at which LLs start to
overlap.

{\it Formalism.} We consider a 2DEG in a classically strong $B$ in
the presence of a weak dc field, $\bE_{dc}=(E_x,E_y)$, and a microwave field
\be
\label{Ew}
\bE_\w(t)=\frac{E_\omega}{\sqrt{2}}\,{\rm Re}\left[\,{s_-
+ s_+\choose i s_-- i s_+}e^{-i\omega t}\,\right]~,
\ee
where $s_\pm$ with $s_+^2+s_-^2=1$ parametrize polarization of the
microwaves.
The main parameters in the problem are related to each other as follows:
\[\ve_F\gg T\,,\,\omega\,,\,\omega_c\,,\,\tau^{-1}_{\rm q} \gg\tau^{-1}_{\rm
tr}\,,\, \tau^{-1}_{\rm in},\]
where $\tau_{\rm q}$ and $\tau_{\rm
tr}$ are the total and transport disorder-induced scattering
times at $B=0$, and $\ve_F$ is the Fermi energy.
We adopt the approach \cite{VA,dmitriev03,long,mechanisms} to the problem,
based on the quantum Boltzmann
equation (QBE) for the semiclassical distribution function at
higher LLs, $f(\ve,\varphi,t)=\sum F_{nm}(\ve)\exp(in\varphi+im\omega
t)$,
\be\label{QBE}
(\partial_t+\wc\partial_\varphi)f+\tau_{\rm in}^{-1}(F_{00}-f_T)=\St\{f\}\,,
\ee
where $\varphi$ is the angle of the kinematic momentum and
$f_T(\ve)$ is the Fermi-Dirac distribution.  The QBE allows us to
treat the interplay of the disorder, the
Landau quantization, and the external fields, which are all
included into the impurity collision integral $\St\{f\}$.

Our aim is to calculate the dissipative dc current, ${\bf
  j}=(j_x,j_y)$, which is expressed through the first angular harmonic
$F_{10}$ as $j_-\equiv j_x-i j_y=2ev_F\int d\ve \nu(\ve) F_{10}(\ve)$,
where $v_F$ is the Fermi velocity. Provided $\tau_{\rm in}\gg\tau_{\rm
  q}$, the leading contribution to the MIRO comes from
microwave-induced changes in the isotropic part $F_{00}$ of the
distribution, governed by the equation $F_{00}(\ve)-f_T(\ve)=\tau_{\rm
  in}\langle\St\{F_{00}(\ve)\}\rangle_{t,\varphi}$. Here the angular
brackets denote averaging over both the angle $\varphi$ and the
period of the microwave field. The first angular harmonic
$F_{10}$, which defines the current, is in turn expressed as $i\wc
F_{10}(\ve)=\langle\exp(-i\varphi)\St\{F_{00}(\ve)\}\rangle_{t,\varphi}$.
As we show below, the above procedure captures both inelastic and
displacement contributions to the MIRO \cite{note-mechanisms},
yielding
\be\label{j}
\frac{j_-}{2\sigma^D}\!=\!\int\!d\Omega\!\int\!d\ve
\,K_{\rm tr}(\Omega)
\tilde\nu(\ve)\tilde\nu(\ve-\Omega)
[F_{00}(\ve-\Omega)-F_{00}(\ve)],\; \ee
\be\label{F00}
\frac{F_{00}(\ve)\!-\!f_T(\ve)}{\tau_{\rm in}}
\!=\!\!\int\!{{d\Omega}\over{2\pi}}K_{00}(\Omega)\tilde\nu(\ve-\Omega)
[F_{00}(\ve-\Omega)-F_{00}(\ve)].
\ee
Here $\sigma^D=e^2 v_F^2\nu_0/2\wc^2\tau_{\rm tr}$ is the Drude
conductivity, 
$\nu_0=m/2\pi$, and
$\tilde\nu(\ve)=\nu(\ve)/\nu_0$
is the dimensionless DOS. The operators $K_{00}(\Omega)$ and $K_{\rm
tr}(\Omega)=K_{10}(\Omega)\,{\wc\tau_{\rm tr}}/{i\pi e v_F}$ can
be found to any desired order in the fields $\bE_{dc}$ and
$\bE_{\w}$ from the Wigner transform ${\cal K}(\Omega, t, \phi)$
of the kernel  of the collision integral $\St\{f\}$ \cite{VA,
mechanisms}, $K_{nm}(\Omega)=\langle e^{ - i n \varphi - i m \w t}%
{\cal K}(\Omega, t,\varphi)\rangle_{t,\varphi}$.
To find the leading contributions to the FMIRO at
half-integer $\w/\wc$, we calculate $K_{nm}$ to first order in
$\bE_{dc}$ and fourth order in $\bE_{\w}$, which gives
\be\label{K}
\left\{{K_{00}(\Omega)\atop K_{\rm tr}(\Omega)}\right\}=\sum\limits^2_{n=-2}\delta(\Omega-n\w)
\left\{{2\pi A_n/\tau_{\rm q}\atop B_n\partial_\Omega}\right\}\,.
\ee
Using the notation $E_\pm=E_x\pm i E_y$ and
\be
{\cal E}_\pm=s_\pm\,({2\tau_{\rm q}}/{\tau_{\rm tr}})^{1/2}\, {e
E_\omega v_F}/{\omega(\wc\pm\omega)}~,
\ee
and introducing the dimensionless microwave power, $P_\w=({\cal E}_+^2+{\cal
E}_-^2)/2$, we express the coefficients
$A_n\!=\!A_{-n}\!\sim\!{\cal O}(P_\w^{n})$ and
$B_n\!=\!B_{-n}\!\sim\!{\cal O}(E_{dc}P_\w^{n})$ as
\bea\nonumber
&&A_1=P_\w/4-4 A_2,\\\nonumber
&&A_2=3P_\w^2/32+3{\cal E}_+^2{\cal E}_-^2/64,\\\nonumber
&&B_0=E_- -2 B_1-2 B_2,\\\nonumber
&&B_1=3P_\w E_-/2 +3{\cal E}_+{\cal E}_-E_+/4   -4 B_2,\\
\label{AB}
&&B_2=45(2P_\w^2 E_- +{\cal E}_+^2{\cal E}_-^2 E_-
+2P_\w{\cal E}_+{\cal E}_- E_+)/64.\qquad
\eea

{\it Integer MIRO.} Before proceeding to the mechanisms of the
FMIRO, it is instructive to show how the results \cite{VA,long} for
the displacement and inelastic contributions to the IMIRO at
leading order $P_\w$ are reproduced within the present
formalism. To this end, we put $A_2=B_2=0$ and calculate $F_{00}$
to first order in $A_1=P_\w/4$, which, according to
Eqs.~(\ref{F00})-(\ref{AB}), gives
\be\label{F1order}F_{00}\!-\!f_T\!=\!{\tau_{\rm in}\over\tau_{\rm q}}
A_1\!\!\sum_{\Omega=\pm\omega}\tilde{\nu}(\ve-\Omega)%
[f_T(\ve-\Omega)-f_T(\ve)].\ee
The result for the current (\ref{j}) at order $P_\w$ has the form
\be\label{1order}
\frac{j_-}{2\sigma^D}\!=\!B_0\langle\tilde{\nu}^2(\ve)\rangle_\ve
\!+\!B_1 {\cal F}_1(\w)\!+\!\frac{\tau_{\rm in}}{\tau_{\rm q}}E_-
A_1 {\cal F}_2(\w), \ee
where the functions ${\cal F}_1(\Omega)$ and ${\cal F}_2(\Omega)$,
defined as
\bea\label{F1} {\cal F}_1(\Omega)&=&2\,\partial_\Omega\,
\Omega\,\langle
\,\tilde{\nu}(\ve)\,\tilde{\nu}(\ve+\Omega)\,\rangle_\ve\,,\\\label{F2}
{\cal
F}_2(\Omega)&=&\Omega\,\partial_\Omega\,\langle\tilde{\nu}^2(\ve)\,
[\,\tilde{\nu}(\ve+\Omega)+\tilde{\nu}(\ve-\Omega)\,]\,\rangle_\ve\,,
\eea
oscillate with $\Omega/\wc$. Here we assumed that the
Shubnikov--de Haas oscillations are exponentially suppressed by
temperature, $2\pi^2 T\gg\wc$, so that the energy integration in
Eq.~(\ref{j}) is effectively replaced by the average
$\langle\ldots\rangle_\ve$ over the period $\wc$ of the DOS.

\begin{figure}[ht]
\centerline{
\includegraphics[width=0.65\columnwidth]{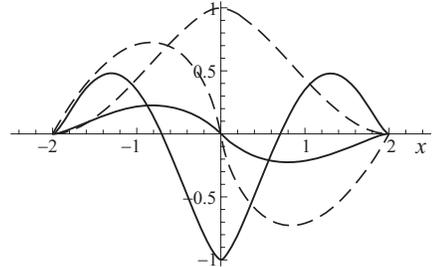}}
\caption{Functions ${\rm sgn}(-x)\Phi_2(|x|)$ and $\Phi_3(|x|)$
(solid lines), ${\cal H}_1(|x|)$ and ${\rm sgn}(x){\cal H}_2(|x|)$
(dashed lines).} \label{fig1}
\end{figure}

In Eq.~(\ref{1order}), the first term describes the dark current
together with a non-oscillatory (displacement) correction induced by
microwaves, while the second and third terms are
the displacement \cite{VA} and inelastic \cite{long} contributions
to the IMIRO, which oscillate with $\w/\wc$.  In the limit of separated
LLs, $\wc\tau_{\rm q}\gg 1$, the DOS is a sequence of semicircles of
width $2\Gamma=2(2\wc/\pi\tau_{\rm q})^{1/2}$, i.e.,
$\tilde{\nu}(\ve)=\tau_{\rm q}{\rm Re}\sqrt{\Gamma^2-(\delta\ve)^2}$,
where $\delta\ve$ is the detuning from the center of the nearest
LL. In this limit, calculation of ${\cal F}_1(\Omega)$ and
${\cal F}_2(\Omega)$ from Eqs.~(\ref{F1}) and (\ref{F2}) yields
\bea\label{f1} &&\frac{{\cal
F}_1(\Omega)}{2\langle\tilde{\nu}^2(\ve)\rangle_\ve}=
\sum\limits_n{\cal H}_1(|\tilde{\Omega}_n|)+
{\Omega\over\Gamma}{\rm sgn}(\tilde{\Omega}_n){\cal
H}_2(|\tilde{\Omega}_n|)\,,\qquad
\\\label{f2}
&&\frac{{\cal F}_2(\Omega)}{\langle\tilde{\nu}^2(\ve)\rangle_\ve}=-{4\Omega\wc\over\Gamma^2}
\sum\limits_n{\rm sgn}(\tilde{\Omega}_n) \Phi_2(|\tilde{\Omega}_n|)\,.
\eea
Here $\tilde{\Omega}_n=(\Omega-n\wc)/\Gamma$ and
$\langle\tilde{\nu}^2(\ve)\rangle_\ve=16\wc/3\pi^2\Gamma$. The
parameterless functions \cite{noteVA} ${\cal H}_1(x)$, ${\cal
H}_2(x)$, and $\Phi_2(x)$ are nonzero at
$0\!<\!x\!<\!2$ (see Fig.~\ref{fig1}),
\bea\label{H1}
{\cal H}_1(x)\!&=&\!(2+x)\,[\,(4+x^2)E(X)-4xK(X)\,]/8\,,\qquad
\\\label{H2}
{\cal H}_2(x)\!&=&\!3x\,[\,(2+x)E(X)-4K(X)\,]/8\,,
\\\label{Phi2}
4\pi\,\Phi_2(x)\!&=&\! 3x\,{\rm arccos}(x-1)-x(1+x)\sqrt{x(2-x)}\,,
\eea
where $X\equiv(2-x)^2/(2+x)^2$ and the functions $E$ and $K$ are the complete elliptic
integrals of the first and second kind, respectively.

{\it Two-photon FMIRO.} At order $E_\w^2$, microwaves do not
produce any oscillatory structure near the fractional values of
$\w/\wc=1/2,\,3/2, \ldots$ [see Eq.~(\ref{1order}]. Moreover, in the
limit $\Gamma\ll\wc$, the oscillatory terms (\ref{1order}) are finite only in
the narrow intervals $(\w-N\wc)<4\Gamma$ around integer $\w/\wc=N$.
Solution to Eqs.~(\ref{F00}), (\ref{j}) at order $E_\w^4$ in the
regions where $\nu(\ve)\nu(\ve+\w)=0$ gives
\be\label{2photon}
\frac{j_-^{(2\phi)}}{2\sigma^D}\!=\!B_0\langle\tilde{\nu}^2(\ve)\rangle_\ve
\!+\!B_2 {\cal F}_1(2\w)\!+\!\frac{\tau_{\rm in}}{\tau_{\rm q}}E_-
A_2 {\cal F}_2(2\w). \ee
Here $A_2\propto P_\w^2$ and $B_2\propto E_{\rm dc} P_\w^2$ are
given by Eq.~(\ref{AB}). The doubling of the arguments of the
functions ${\cal F}_1$ and ${\cal F}_2$ in Eq.~(\ref{2photon}) [as
compared to the IMIRO case, Eq.~(\ref{1order})] reflects the
two-photon nature of the effect and leads to the emergence of the
FMIRO at half-integer $\w/\wc$. The form of the inelastic [the
last term in Eq.~(\ref{2photon})] contribution to the FMIRO is identical
to that in the IMIRO case, Eq.~(\ref{1order}), and the same holds
for the displacement contribution (the second term). In both the
integer and fractional cases the inelastic term is a factor
$\wc\tau_{\rm in}/\tau_{\rm q}\Gamma$ larger than the displacement
contribution.

  With increasing microwave power, 
  the current in the minima becomes
  negative, ${\bf j\cdot E_{\rm dc}}<0$, indicating a transition
  to the ZRS \cite{andreev03}. Remarkably, like in the IMIRO case \cite{long},
  the leading-order approximation for the inelastic  effect,
  Eq.~(\ref{2photon}), is sufficient to describe the photoresponce
  even at such high power, since the second-order term $\propto (P_\w^2\tau_{\rm
  in}/\tau_{\rm q})^2$ remains small in the parameter $\Gamma/\wc$.

\begin{figure}[ht]
\centerline{
\includegraphics[width=0.7\columnwidth]{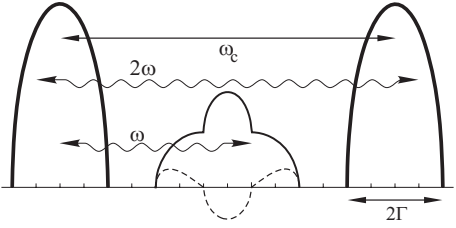}
} \caption{ Illustration of possible contributions to the FMIRO
for $\w/\wc=1/2+\Gamma/2$ and $\wc/\Gamma=7$.
Single-photon transitions within and between LLs (thick lines) are
forbidden, while two-photon processes are allowed. The microwave-induced
sidebands (\ref{nusb}) (solid lines) and
(\ref{gosb}) (dashed lines) make
single-photon processes possible \cite{confusion}. } \label{fig2}
\end{figure}

{\it Microwave-induced spectral reconstruction (MISR).} Above, it
was assumed that the spectrum of the 2DEG is not modified by the
microwaves. It turns out that, unlike the IMIRO, in the FMIRO case
the MISR is relevant, since it provides additional channels for
electron transport in the gaps for single-photon absorption.
The microwave field enters the
equations for the spectrum of the 2DEG at high LLs \cite{VA} in
the combination $[\sum_\pm{\cal E}_\pm\cos(\varphi\pm\w t)]^2$,
suggesting a representation of the
retarded Green function and self-energy in the form
$\{G_{\hat{n}},\Sigma\}(\ve,\hat{\varphi},t)= \sum_{\mu,\nu}e^{2
i\mu\omega t+2 i\nu\hat{\varphi}}
\{G_{\hat{n}\mu\nu},\Sigma_{\mu\nu}\}(\ve)$,
where the operators $\hat{n}$ and $\hat{\varphi}$ obey
$[\hat{n},\hat{\varphi}]=-i$. Wigner-transforming Eqs.~(2.7),
(2.15) of Ref.~\onlinecite{mechanisms} yields
\bea\nonumber
[\ve-\mu\w-(2\nu+n+1/2)\wc]G_{n\mu\nu}(\ve)
=\delta_{\mu,0}\delta_{\nu,0}/2\pi\\\label{dyn}
+\sum_{\mu', \nu'}\Sigma_{\mu' \nu'}(\ve+\mu'\w-\mu\w)
G_{n,\mu-\mu', \nu-\nu'}(\ve+\mu'\w),\\\nonumber
{\wc\over\tau_{\rm q}}\sum_n G_{n\mu\nu}(\ve)=
\Sigma_{\mu\nu}(\ve)+\sum_{\mu', \nu'}p_{\mu-\mu', \nu-\nu'}\phantom{aaaaaa}\\\label{SCBA}
\times[2\Sigma_{\mu'\nu'}(\ve)-\Sigma_{\mu'\nu'}(\ve+\w)-\Sigma_{\mu'\nu'}(\ve-\w)],
\eea
where we used
$\hat{n}e^{2i\nu\hat{\varphi}}=e^{2i\nu\hat{\varphi}}(\hat{n}+2\nu)$.
The coefficients $p_{\mu\nu}=p_{-\mu,-\nu}$ have nonzero values
for $\{\mu,\nu\}=\{-1,0,1\}$: $p_{00}=P_\w/4$,
$p_{10}=p_{01}={\cal E}_+{\cal E}_-/8$, $p_{11}={\cal E}_+^2$,
$p_{1-1}={\cal E}_-^2$.

Solution to Eqs.~(\ref{dyn}), (\ref{SCBA}) of order $E_\w^2$ for
$g_{\mu\nu}(\ve)\equiv -2\wc\sum_n {\rm
  Im}G^{(1)}_{n\mu\nu}(\ve)$, which enters Eq.~(\ref{kineq})
  for $F_{00}$, is expressed through the zero-order
self-energy $s(\ve)\equiv\Sigma^{(0)}_{00}(\ve)$ and $2\pi
G^{(0)}_n(\ve)=[\ve_n-s(\ve)]^{-1}$:
\be\label{MISR}
g^{(1)}_{\mu\nu}(\ve)=-2 p_{\mu\nu}{\rm Im}
\frac{s(\ve+\w)+s(\ve-\w)-2s(\ve)}
{\Pi_{\mu\nu}^{-1}-\tau_{\rm q}^{-1}}~,
\ee
where $\Pi_{\mu\nu}(\ve)=2\pi\wc\sum_n
G^{(0)}_{n+2\nu}(\ve-\mu\w)G^{(0)}_n(\ve+\mu\w)$ and
$\ve_n\equiv\ve-n\wc-\wc/2$.  In the limit $\wc\tau_{\rm q}\gg 1$,
the self-energy $s(\ve)={1\over2}\sum_n\ve_n-{\rm
  Re}\sqrt{\ve_n^2-\Gamma^2}-i{\rm Re}\sqrt{\Gamma^2-\ve_n^2}$.


{\it Sideband mechanism of the FMIRO.} The MISR, Eq.~(\ref{MISR}),
gives rise to additional, single-photon contributions to the FMIRO
near half-integer $\w/\wc$ (see Fig.~\ref{fig2}). Restricting the
following analysis to the inelastic effects, we rewrite
Eq.~(\ref{F00}) in a generalized form:
\bea\nonumber &&F_{00}(\ve)-f_T(\ve)=\tau_{\rm in}
\sum_{\mu\nu}\int\! {{d\Omega}\over {2\pi}}
K_{-2\mu,-2\nu}(\Omega)\\\label{kineq} &&\times
\,g_{\mu\nu}(\ve-\Omega) [F_{00}(\ve-\Omega+\mu\w)-F_{00}(\ve)],
\eea
which takes the form of Eq.~(\ref{F00}) in the case of a $t$ and
$\varphi$ independent spectrum; in particular, for the unperturbed
DOS, $\tilde{\nu}(\ve)=g^{(0)}_{00}=-2\tau_{\rm q}\,{\rm
Im}~s(\ve)$.  At order $E_\w^2$, $\tau_{\rm
q}K_{2\mu,2\nu}(\Omega)=2\pi
p_{\mu\nu}[\delta(\Omega+\w)+\delta(\Omega-\w)-2\delta(\Omega)]$.

According to Eq.~(\ref{MISR}), $g^{(1)}_{\mu\nu}$ have nonzero
values at energies inside the unperturbed LLs
($|\delta\ve|<\Gamma$), where $\tilde{\nu}(\ve)\neq 0$, and also
at $\ve$ corresponding to $\tilde{\nu}(\ve\pm\omega)\neq0$
(``microwave-induced sidebands''). Provided $\w=(N+1/2)\wc+{\cal
O}(\Gamma)$, the sidebands appear at $|\delta\ve|=\wc/2+{\cal
O}(\Gamma)$, where Eq.~(\ref{MISR}) gives, to leading order in
$\Gamma/\wc$ \cite{confusion,01}:
\be\label{nusb}
\tilde{\nu}^{(sb)}(\ve)=g^{(1)}_{00}=
\frac{\pi p_{00}}{2\wc\tau_{\rm
q}}[\tilde{\nu}(\ve+\w)+\tilde{\nu}(\ve-\w)]~,
\ee
\vspace*{-4.5mm}
\be\label{gosb}
g^{(1)}_{\mu=\pm1,\nu}(\ve)\!=\!\frac{\tau_{\rm q}
p_{\mu\nu}}{\mu\w+\nu\wc}{\rm Im}[s^2(\ve\!+\!\w)\!-\!s^2(\ve\!-\!\w)].
\ee

In the presence of the sidebands, single-photon transitions become
possible (Fig.~\ref{fig2}),
$\tilde{\nu}(\ve)g_{\mu\nu}^{(1)}(\ve\pm\w)\neq0$.  A correction
to $F_{00}$, induced by the sidebands in the DOS (\ref{nusb}), is
given by Eq.~(\ref{F1order}) with $\tilde{\nu}$ substituted by
$\tilde{\nu}^{(sb)}$.  Calculation of the current (\ref{j}) with
the resulting $F_{00}$, the unperturbed DOS, and with $K_{\rm
tr}=\delta(\Omega)\,E_-\,\partial_\Omega$ yields the ``sideband''
contribution to the FMIRO:
\be\label{sb}{j_-^{(sb)}}/{2\sigma^D}=\pi E_-\tau_{\rm
in}P_\w^2{\cal
F}_2(2\w)/64\wc\tau_{\rm q}^2\,, \ee
which is of the same form but a factor $\sim\wc\tau_{\rm q}$
smaller than the leading two-photon inelastic contribution
(\ref{2photon}).

For the ``oscillatory sidebands'' (\ref{gosb}), we use
Eq.~(\ref{kineq}) to obtain $F_{00}$, which yields the current
\be \label{osb} \frac{j_-^{(osb)}}{2\sigma^D}=\frac{\tau_{\rm
in}}{16\tau_{\rm q}}E_- \left[4{\cal E}_+^2{\cal
E}_-^2+\sum_\pm\frac{\w{\cal E}_\pm^4}{\w\pm\wc} \right] {\cal
F}_3(2\w). \ee
Here ${\cal F}_3(\Omega)$ is an even function of
$\tilde{\Omega}_n=(\Omega-n\wc)/\Gamma$:
\bea\nonumber &&{\cal F}_3(\Omega)\!=\!\sum_\pm\langle\,
[\tilde{\nu}^2(\ve)-\tilde{\nu}^2(\ve\pm\Omega)]\,\partial_\ve\,
[\tilde{\nu}(\ve){\rm Re}~s(\ve)]\,\rangle_\ve\\\label{phi3}
&&=\frac{2\wc}{3\Gamma}\langle\tilde{\nu}^2(\ve)\rangle_\ve
\left[1+\sum\limits_n
\Phi_3(|\tilde{\Omega}_n|)\,\theta(2-|\tilde{\Omega}_n|)\right]\,.
\eea
The parameterless function $\Phi_3(x)$ (Fig.~\ref{fig1}) reads
\[
-\pi\,\Phi_3(x)\!=\!\sqrt{x(2-x)}\left[1\!-\!5x\!+\!2x^2\frac{1+x}{3}\right]\!+\!{\rm arccos}(x\!-\!1).
\]
The contribution of the oscillatory sidebands (\ref{osb}) is a factor
$\sim(\wc\tau_{\rm q})^{1/2}$ larger than the contribution
(\ref{sb}). In contrast to the contributions (\ref{2photon}) and
(\ref{sb}), it is an even and strongly non-monotonic function of
the detuning from the resonance (see Fig.~\ref{fig1}). Both
these circumstances favor the possibility of observing experimentally the
changes in the shape of the FMIRO that should be induced by the contribution (\ref{osb})
at not too large $\wc\tau_{\rm q}$.

 {\it Summary and disscussion.} We have shown that in
the limit of separated LLs, $\wc\tau_{\rm q}\gg 1$, the fractional
features in the photoresponse of a 2DEG are dominated by the
multiphoton inelastic mechanism [the last term in
Eq.~(\ref{2photon})], while the displacement multiphoton
contribution [the second term in Eq.~(\ref{2photon})], considered
in Ref.~\onlinecite{leiliumulti}, is negligible provided
$\wc\tau_{\rm in}/\tau_{\rm  q}\Gamma\gg1$.
  The main corrections (\ref{sb}), (\ref{osb}) to the multiphoton
  inelastic effect originate from the microwave-induced sidebands (\ref{nusb}), (\ref{gosb})
  in the spectrum of a 2DEG (see Fig.~\ref{fig2}).
  In the limit $\wc\tau_{\rm q}\gg 1$, the sideband contributions are small in
the parameter $\Gamma/\wc$
  and only slightly modify the shape and the amplitude of the FMIRO, which are dominated by the multiphoton
inelastic mechanism.
  However,  at $\wc\tau_{\rm q}\sim 1$ all three inelastic contributions
(\ref{2photon}), (\ref{sb}), and (\ref{osb}) become comparable in magnitude.

At $\wc<4\Gamma$, single-photon processes both within and between LLs
become allowed. Here, the FMIRO are dominated by the resonant
series of multiple single-photon transitions
\cite{dorozhkin06,crossover} in the framework of the
single-photon inelastic mechanism \cite{long}. Albeit this effect only exists
in the crossover region $\wc\sim\Gamma$, it appears at order $(\tau_{\rm in}
P_\w/\tau_{\rm q})^2$ and is thus detectable at smaller
$P_\w$ than the above contributions. Finally, in the regime of
overlapping LLs, $\wc\tau_{\rm q}\ll 1$, the FMIRO become
exponentially weaker than the IMIRO: while the $B$-damping of the
IMIRO is described by the factor $\delta^2=\exp(-2\pi/\wc
\tau_{\rm q})\ll 1$ \cite{long}, the FMIRO appear at order
$\delta^4$ \cite{crossover,mechanisms}.

To conclude, there are competing mechanisms of the
fractional oscillations, each of which is effective in a different
region of magnetic field. The experimental observations \cite{dorozhkin06}
should be attributed to the single-photon inelastic mechanism
\cite{crossover,mechanisms} since the FMIRO in Ref.~\cite{dorozhkin06} were only
observed for microwave frequencies below a certain threshold. By
contrast, in Ref.~\cite{multi} no frequency threshold was reported and
strongly developed ZRS were observed, which favors the explanation
in terms of the mechanisms we consider here. It would
be of interest to perform experiments on the FMIRO in the regime
of well-separated LLs, where gaps in the single-photon absorption
spectrum were observed \cite{dorozhkinINTRA}. In particular, to
distinguish between the different mechanisms, we suggest to
measure and compare the power and temperature dependencies of the
photoresponse at $\Gamma\ll\wc$ and $\Gamma\sim\wc$.

We thank S.I.Dorozhkin and
M.A.~Zudov for information about the experiments,
 and I.V.~Gornyi for
discussions. This work was
supported by INTAS Grant
No.~05-1000008-8044, by the DFG-CFN, and by the RFBR.


\begin{thebibliography}{1}
\bibitem{note-mechanisms} As far as the MIRO in the Hall
resistivity are not concerned and $\w_c\tau_q\gg 1$ ,
contributions to the photocurrent involving higher temporal and angular harmonics $F_{nm}$
(different from $F_{00}$ and $F_{10}$) can be neglected \cite{mechanisms}.
\bibitem{confusion} To avoid confusion, the ``oscillatory
sidebands'' (\ref{gosb}) have nothing to do with the DOS given by $g_{00}$.
\bibitem{01}The
contribution of $g^{(1)}_{0,\pm1}$ is neglected since
$\Pi_{0,\pm1}\sim{\cal O}(\wc/\ve_F)$.
\end{thebibliography}
\end{document}